\documentclass{article}
\usepackage{spconf}

\usepackage{amsmath,amssymb,amsfonts}
\usepackage{amsthm}

\usepackage{graphicx}
\usepackage{subcaption}
\usepackage{booktabs}
\usepackage{multirow}
\usepackage{caption}

\usepackage{algorithm}
\usepackage{algorithmic}

\usepackage{xcolor}
\usepackage{nicefrac}
\usepackage{microtype}
\usepackage{url} % (keep only one)

\newtheorem{lemma}{Lemma}[section]
\newtheorem{corollary}[lemma]{Corollary}

\usepackage{hyperref}

\newcommand{\X}{\mathbf{X}}
\newcommand{\Q}{\mathbf{Q}}

\newcommand{\K}{\mathbf{K}}
\newcommand{\V}{\mathbf{V}}
\newcommand{\G}{\mathbf{G}}
\newcommand{\W}{\mathbf{W}}
\newcommand{\M}{\mathbf{M}}
\newcommand{\I}{\mathbf{I}}
\newcommand{\boldO}{\mathbf{O}}
\newcommand{\A}{\mathbf{A}}
\renewcommand{\S}{\mathbf{S}}

\title{Plug-and-Play Linear Attention with Provable Guarantees for Training-Free Image Restoration}

\name{Srinivasan Kidambi$^{\star}$ \qquad Karthik Palaniappan$^{\star}$ \qquad Pravin Nair$^{\star}$}

\address{
$^{\star}$Indian Institute of Technology, Madras}
\begin{document}
%\ninept
%
\maketitle

\begin{abstract}
Multi-head self-attention (MHSA) is a key building block in modern vision Transformers, yet its quadratic complexity in the number of tokens remains a major bottleneck for real-time and resource-constrained deployment. We present {PnP-Nystra}, a {training-free} Nystr\"{o}m-based {linear attention} module designed as a plug-and-play replacement for MHSA in {pretrained} image restoration Transformers, with provable kernel approximation error guarantees. PnP-Nystra integrates directly into window-based architectures such as SwinIR, Uformer, and Dehazeformer, yielding efficient inference without finetuning. Across denoising, deblurring, dehazing, and super-resolution on images, PnP-Nystra delivers $1,8$--$3.6\times$ speedups on an NVIDIA RTX 4090 GPU and $1.8$--$7\times$ speedups on CPU inference. Compared with the strongest training-free linear-attention baselines we evaluate, our method incurs the smallest quality drop and stays closest to the original model's outputs.
\end{abstract}

\section{Introduction}
\label{sec:intro}

Transformers have recently demonstrated strong performance across a range of low-level vision tasks, including image denoising, deblurring~\cite{uformer}, super-resolution~\cite{swinir}, and dehazing~\cite{song2023vision}.  These models rely on a self-attention mechanism that captures long-range dependencies by computing pairwise similarity between features corresponding to image patches. Let $\X \in \mathbb{R}^{N \times d_i}$ be the input feature map corresponding to $N$ tokens (e.g., pixels or patches), where each token is represented by a $d_i$-dimensional feature vector. To compute self-attention \cite{attn}, the input $\X$ is first projected into query, key, and value representations:
\begin{equation*}
\Q = \X \W_Q, \quad 
\K = \X \W_K, \quad 
\V = \X \W_V,
\end{equation*}
where $\W_Q, \W_K \in \mathbb{R}^{d_i \times d}$, $\W_V \in \mathbb{R}^{d _i\times d_v}$ are learned projection matrices. Then, the attention matrix $\A \in \mathbb{R}^{N \times N}$, its row-normalized form $\S \in \mathbb{R}^{N \times N}$, and the self-attention output $\boldO \in \mathbb{R}^{N \times d_v}$ are computed as:
\begin{equation}
\A = \frac{\Q \K^\top}{\sqrt{d}}, \qquad 
\S = \operatorname{softmax}(\A), \qquad 
\boldO = \S \V,
\label{eq:selfattention}
\end{equation}
where $\operatorname{softmax}(\A)$ denotes independently applying softmax to each row of $\A$. 

However, the attention operation in \eqref{eq:selfattention} incurs a quadratic complexity in the number of input tokens \( N \), which corresponds to the number of pixels or patches within a subregion of a frame~\cite{attn}. In particular, 
the time and storage complexity for computing \eqref{eq:selfattention} is $\mathcal{O}(N^2 d)$ and $\mathcal{O}(N^2 + N d)$, respectively, dominated by storing and computing the large attention matrices $\A$ and $\S$. Currently, most vision models employ Multi-Head Self-Attention (MHSA) \cite{attn}, an extension of self-attention. In MHSA, the embedding dimension $d$ is divided into $h$ attention heads, each of dimension $d_h$, such that $d = h d_h$. For each head, self-attention outputs are independently computed, concatenated, and linearly projected to form the MHSA output. The time and storage complexity of MHSA is quadratic in $N$, similar to self-attention. This computational bottleneck makes transformers less suitable for real-time or resource-constrained settings, where convolutional networks are still often preferred due to their computational efficiency~\cite{convnet,ren2024ninth}.

Several approaches approximate self-attention via low-rank~\cite{linformer,choromanski2020rethinking,xiong2021nystromformer} or sparse~\cite{sparseformer} representations to reduce quadratic complexity, but these methods often require retraining. Linformer~\cite{linformer} projects keys/values into lower-dimensional subspaces via learned linear maps, achieving linear complexity in \(N\). Performer~\cite{choromanski2020rethinking} replaces the softmax kernel with random feature approximation, resulting in linear complexity and unbiased attention estimates. Nystr\"{o}mformer~\cite{xiong2021nystromformer} heuristically reconstructs attention using landmark-based sampling. Axial Attention~\cite{ho2019axial} factorizes 2D attention into 1D along spatial axes, reducing cost from \(H^2W^2\) to \(HW(H+W)\), where the number of tokens is $N=HW$. FlashAttention~\cite{flashattention} and Triton~\cite{tillet2019triton} preserve exact attention while optimizing execution through custom fused GPU kernels, but are hardware-specific and difficult to integrate into pre-trained models.

Interestingly, research on low-rank attention has seen reduced momentum since 2022, with the focus shifting to hardware-aware accelerations. In this work, we revisit classical low-rank approximations that can act as a {plug-and-play} replacement for attention modules in transformer-based vision models. Specifically, we propose \textbf{PnP-Nystra}, a training-free, plug-and-play approximation of self-attention using the Generalized Nystr\"{o}m approximation~\cite{nemtsov2016matrix, mahoney2009cur, wang2013improving}. 
The core contributions of our method are summarized as follows:

\noindent \textbf{1) Linear attention:} Unlike MHSA, which incurs quadratic time and memory complexity in the number of tokens \( N \), {PnP-Nystra} achieves linear scaling in both time and memory.

\noindent \textbf{2) Impressive acceleration:} PnP-Nystra serves as a drop-in replacement for standard self-attention in existing window-attention transformer backbones for image restoration, enabling substantial inference-time acceleration without modifying the pretrained weights. In our experiments, the GPU inference speedup ranges from $1.8\times$ to $3.6\times$ across restoration tasks, while the gains are even larger on CPU inference.

\noindent \textbf{3) Robust performance without retraining:} Despite significant acceleration, the reduction in restoration quality is minimal. {For instance, in image denoising, PnP-Nystra achieves more than $3.5\times$ speedup on CPU with $<1$ dB drop in PSNR and $<0.03$ drop in SSIM.}

To the best of our knowledge, we are the first to show that a purely drop-in linear-attention replacement can remain close to the original pretrained self-attention outputs without any finetuning, while retaining provable approximation guarantees.

\section{Proposed Method}
\label{sec:method}

In \eqref{eq:selfattention}, by letting \( \mathbf{G} = \exp(\mathbf{A}) \) denote the element-wise exponential of the attention matrix \( \mathbf{A} \), the self-attention output \( \mathbf{O} \in \mathbb{R}^{N \times d_v} \) can be reformulated as
\begin{equation}
\mathbf{O} = \left( \mathbf{G} \mathbf{V} \right) \oslash \left( \mathbf{G} \mathbf{1}_N \right),
\label{eq:compact_attention}
\end{equation}
where \( \mathbf{1}_N \in \mathbb{R}^{N} \) is the all-ones column vector, and \( \oslash \) denotes row-wise element-wise division: that is, the \( i \)-th row of \( \mathbf{G} \mathbf{V} \) is divided by the scalar \( (\mathbf{G} \mathbf{1}_N)_i \). Thus, the attention output \( \mathbf{O} \in \mathbb{R}^{N \times d_v} \) can be expressed in the kernel-based form from~\eqref{eq:compact_attention}. Specifically, let the query matrix \( \mathbf{Q} \in \mathbb{R}^{N \times d} \) and key matrix \( \mathbf{K} \in \mathbb{R}^{N \times d} \) be viewed as stacks of row vectors \( \{ \boldsymbol{q}_i \}_{i=1}^N \) and \( \{ \boldsymbol{k}_j \}_{j=1}^N \), respectively. Then, the kernel matrix \( \mathbf{G} \) in~\eqref{eq:compact_attention} is defined element-wise, using an exponential kernel, as
\[
\mathbf{G}_{ij} = \exp\left( \frac{\boldsymbol{q}_i^\top \boldsymbol{k}_j}{\sqrt{d}} \right).
\]
We propose to leverage the low-rank structure of the kernel matrix $\mathbf{G}$ via a generalized Nystr\"{o}m method adapted for non-symmetric matrices. Originally developed for approximating solutions to functional eigenvalue problems \cite{nystrom1930praktische, baker1977numerical}, the Nystr\"{o}m method was later extended to efficiently estimate eigenvectors and construct low-rank matrix approximations \cite{talebi2014global,williams2000using, fowlkes2004spectral}. 
To apply the Nystr\"{o}m approximation, we first select $m \ll N$ landmark query and key vectors, denoted by $\{\boldsymbol{\bar{q}}_i\}_{i=1}^m$ and $\{\boldsymbol{\bar{k}}_i\}_{i=1}^m$. Let $\bar{\Q} \in \mathbb{R}^{m \times d}$ and $\bar{\K} \in \mathbb{R}^{m \times d}$ denote the matrices formed by stacking these landmark vectors. We define the extended query and key matrices:
\[
\widetilde{\Q} = 
\begin{bmatrix}
\bar{\Q} \\ \Q
\end{bmatrix}, \qquad
\widetilde{\K} = 
\begin{bmatrix}
\bar{\K} \\ \K
\end{bmatrix}.
\]
We then define the extended kernel matrix $\widetilde{\G} \in \mathbb{R}^{(m+N) \times (m+N)}$ as:
\begin{equation}
\widetilde{\G} = \exp\left( \frac{\widetilde{\Q} \widetilde{\K}^\top}{\sqrt{d}} \right) = 
\begin{bmatrix}
\G_A & \G_U \\
\G_L & \G
\end{bmatrix},
\label{eq:gtilde}
\end{equation}
where, the core matrix $\G_A \in \mathbb{R}^{m \times m}$ captures similarities between landmark queries and keys:
\begin{equation}
\G_A = \exp\left(\frac{\bar{\Q} \bar{\K}^\top}{\sqrt{d}}\right).
\label{landmarkmatrix}
\end{equation}
The cross-similarity matrices are defined as $\G_L \in \mathbb{R}^{N \times m}$ (queries with landmark keys) and $\G_U \in \mathbb{R}^{m \times N}$ (landmark queries with keys):
\begin{equation}
\G_L = \exp\left(\frac{\Q \bar{\K}^\top}{\sqrt{d}}\right), \quad
\G_U = \exp\left(\frac{\bar{\Q} \K^\top}{\sqrt{d}}\right).
\label{crossmatrix}
\end{equation}
\noindent Applying Nystr\"{o}m approximation to the extended matrix $\widetilde{\G}$, the original kernel matrix $\G$ is approximated as:
\begin{equation}
\widehat{\G} = \G_L \G_A^\dagger \G_U,
\label{eq:Nystromapprox}
\end{equation}
where $\G_A^\dagger$ denotes the Moore-Penrose pseudoinverse. Substituting \eqref{eq:Nystromapprox} into \eqref{eq:compact_attention}, the attention output is approximated as:
\begin{equation}
\widehat{\boldO} = \left( \widehat{\G} \V \right) \oslash  \left( \widehat{\G} \mathbf{1}_N \right).
\label{eq:compact_approxattention}
\end{equation}
We refer to the approximation in~\eqref{eq:compact_approxattention} as \textbf{PnP-Nystra}, our method for approximating self-attention. The complete steps for computing PnP-Nystra are summarized in Algorithm~\ref{alg:pnpnystra}.

\begin{algorithm}[h]
\caption{PnP-Nystra in \eqref{eq:compact_approxattention}: Self-Attention Approximation}
\label{alg:pnpnystra}
\begin{algorithmic}
\REQUIRE Queries \( \Q \in \mathbb{R}^{N \times d} \), Keys \( \K \in \mathbb{R}^{N \times d} \), Values \( \V \in \mathbb{R}^{N \times d_v} \), number of landmarks \( m \)

\STATE \textbf{(Step 1)} Select \( m \) landmark query and key vectors to form landmark matrices \( \bar{\Q} \in \mathbb{R}^{m \times d}, \bar{\K} \in \mathbb{R}^{m \times d} \)

\STATE \textbf{(Step 2)} Compute core matrix \(\G_A \in\mathbb{R}^{m \times m}\)
in \eqref{landmarkmatrix}

\STATE \textbf{(Step \hspace{-0.07cm}3)} Compute cross-similarity matrices \( \G_L \in \mathbb{R}^{N \times m} \) and  \( \G_U \in \mathbb{R}^{m \times N} \) in  \eqref{crossmatrix}
 
\STATE \textbf{(Step 4)} Compute pseudoinverse \( \G_A^\dagger \in \mathbb{R}^{m \times m} \)

\STATE \textbf{(Step 5)} Compute premultiplier \( \mathbf{P} \in \mathbb{R}^{N \times m}$ as $\G_L \G_A^\dagger  \)

\STATE \textbf{(Step 6)} Compute numerator \( \boldO_N = \mathbf{P} \, (\G_U \V) \in \mathbb{R}^{N \times d_v} \)

and compute denominator \( \boldO_D = \mathbf{P} \, (\G_U \mathbf{1}_N) \in \mathbb{R}^{N \times 1} \)

\STATE \textbf{(Step 7)} Normalize \( \widehat{\boldO} = \boldO_N \oslash \boldO_D \) \hfill \textit{\small (row-wise division)}

\RETURN Output \( \widehat{\boldO} \in \mathbb{R}^{N \times d_v} \)
\end{algorithmic}
\end{algorithm}

Note that, while the Nystr\"{o}mformer \cite{xiong2021nystromformer} also employs the Nystr\"{o}m method, it operates on the softmax-transformed attention matrix $\mathbf{S}$. However, the softmax operation prevents $\mathbf{S}$ from being naturally partitioned into submatrices - a core requirement for Nystr\"{o}m-based approximation. As a result, the authors resort to a heuristic strategy, which makes their approximation devoid of  guarantees such as those in Lemma~\ref{lem:Nystrom_lemma}.

\subsection{Approximation error}
\begin{figure}
  \centering
  \begin{subfigure}[b]{0.48\columnwidth}
    \centering
     \includegraphics[width=\textwidth]{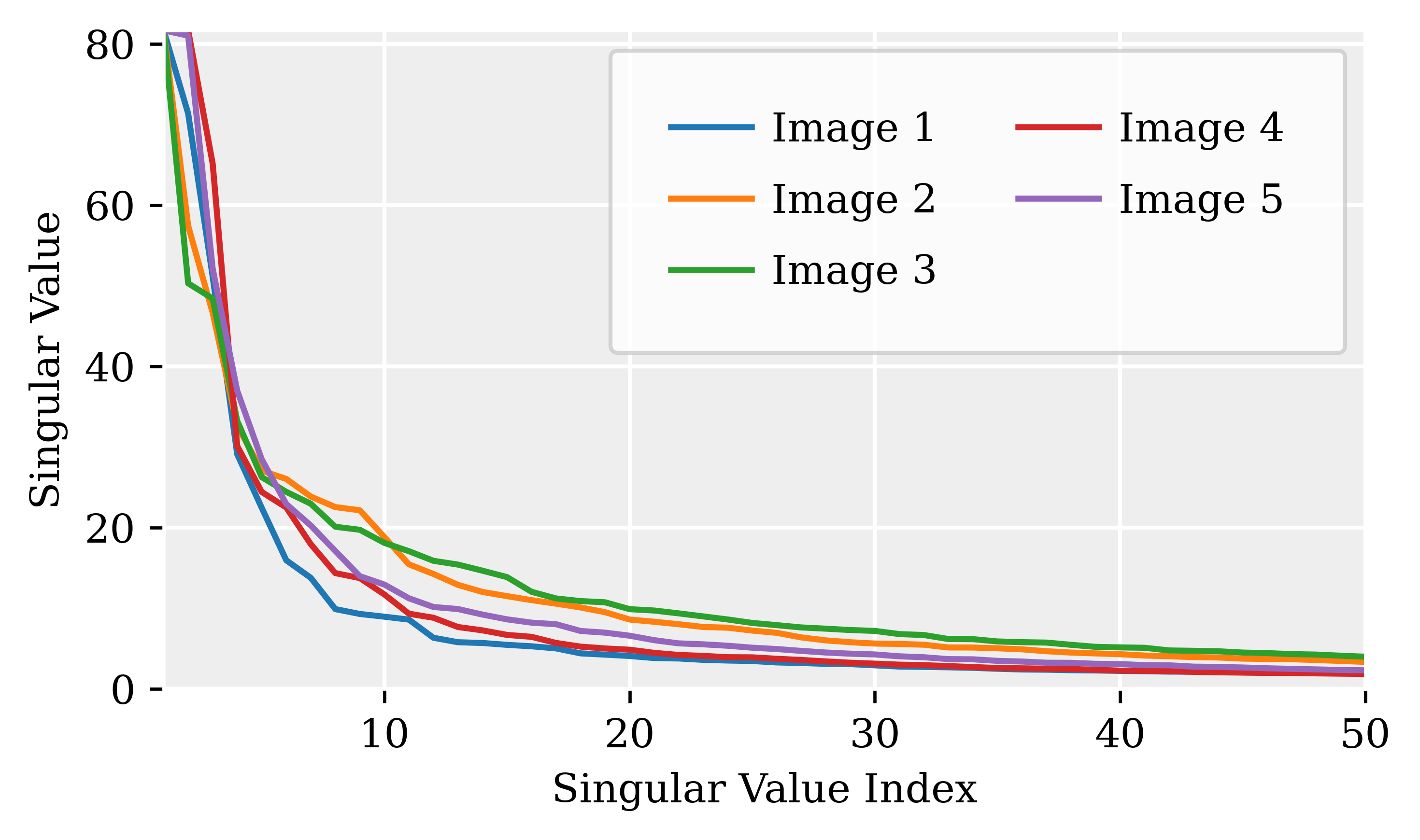}
    \caption{SwinIR \cite{swinir}}
    \label{fig:batchheadavg}
  \end{subfigure}\hfill
  \begin{subfigure}[b]{0.48\columnwidth}
    \centering
    \includegraphics[width=\textwidth]{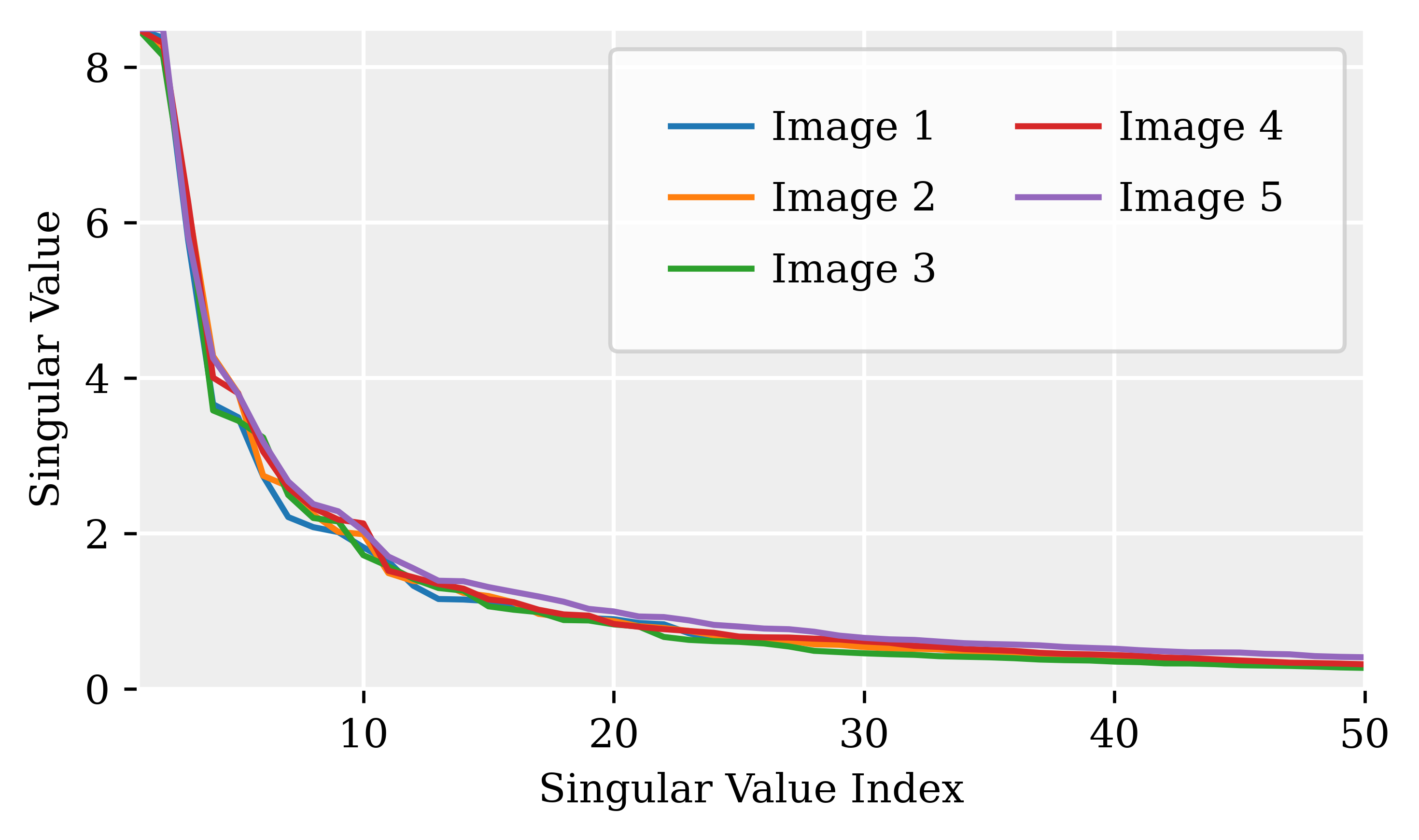}
    \caption{Uformer-B \cite{uformer}}
    \label{fig:bacthavg}
  \end{subfigure}
   \caption{Variation of the top $50$ singular values of attention maps ($N = 32^2$) from SwinIR and Uformer, averaged over all heads and layers. The steep decay within the first $20$ singular values highlights the low-rank structure of the attention matrices.
 }
  \label{fig:singulars}
\end{figure}

We analyze the spectral norm error of the Nystr\"{o}m-based kernel approximation used in PnP-Nystra, where the original matrix \( \mathbf{G} \) is approximated as \( \widehat{\mathbf{G}} = \mathbf{G}_L \mathbf{G}_A^\dagger \mathbf{G}_U \).
\begin{lemma}[Spectral norm error bound] \label{lem:Nystrom_lemma}
Assume \( \mathbf{G}_A \) is nonsingular. Then,
\begin{equation*}
\| \mathbf{G} - \mathbf{G}_L \mathbf{G}_A^\dagger \mathbf{G}_U \|_2 
\le C \, \frac{\sigma_{m+1}(\widetilde{\mathbf{G}})}{\sigma_m(\mathbf{G}_A)},
\end{equation*}
where \( \sigma_k(\cdot) \) denotes the \( k \)-th singular value, and the constant \( C \) depends quadratically on \( \sigma_1(\widetilde{\mathbf{G}}) \).
\end{lemma}
The proof follows from Lemma~5.10 in~\cite{nemtsov2016matrix}.
% , along with the  fact that the spectral norm error between \( \mathbf{G} \) and \( \widehat{\mathbf{G}} \)  is upper bounded by that between \( \widetilde{\mathbf{G}} \) and its Nystr\"{o}m approximation ($\widehat{\G}$ instead of $\G$ in \eqref{eq:gtilde}). 
The error bound in Lemma \ref{lem:Nystrom_lemma} indicates that the approximation error decreases when the singular values of \( \widetilde{\mathbf{G}} \) decay rapidly. In particular, if the number of landmark points \( m \) approaches the rank of \( \widetilde{\mathbf{G}} \), then \( \sigma_{m+1}(\widetilde{\mathbf{G}}) \to 0 \), leading to tighter approximations. The following result formalizes this by guaranteeing perfect reconstruction when the number of landmark points equals the rank of the extended kernel matrix. 
\begin{corollary}[Exact recovery]\label{exactNystrom}
If \( \operatorname{rank}(\widetilde{\mathbf{G}}) = m \), then \( \sigma_{m+1}(\widetilde{\mathbf{G}}) = 0 \), and the Nystr\"{o}m approximation becomes exact, i.e., \( \mathbf{G} = \mathbf{G}_L \mathbf{G}_A^\dagger \mathbf{G}_U \).
%\label{exactNystrom}
\end{corollary}
We next empirically validate the low-rank property of the kernel matrix in pre-trained vision models, which is indeed the assumption underlying Lemma~\ref{lem:Nystrom_lemma} and Corollary~\ref{exactNystrom}. Fig.~\ref{fig:singulars} illustrates the singular value spectrum of the attention matrices from two distinct pre-trained models. In both cases, the rapid singular value decay confirms the low-rank structure of the attention matrices.

\subsection{Computational Details}

Standard self-attention has time complexity $\mathcal{O}(N^2d)$ and memory $\mathcal{O}(N^2+Nd)$. By avoiding explicit $N\times N$ attention, PnP-Nystra reduces the overall complexity to $\mathcal{O}(Nm(d+d_v))$ time and $\mathcal{O}(m^2+N(m+d_v))$ memory for $m\ll N$ (Table~\ref{tab:complexities}). 
\begin{table}[!t]
  \centering
  \caption{Per-step time/space complexity of PnP-Nystra.}
  \label{tab:complexities}
  \small
  \resizebox{\columnwidth}{!}{
  \begin{tabular}{l|cccccc}
    \toprule
    \textbf{Step} & 2 & 3 & 4 & 5 & 6 & 7 \\ \midrule
    Time  & $\mathcal{O}(m^2d)$ & $\mathcal{O}(Nmd)$ & $\mathcal{O}(m^3)$ & $\mathcal{O}(Nm^2)$ & $\mathcal{O}(Nm(d_v{+}1))$ & $\mathcal{O}(Nd_v)$ \\
    Space & $\mathcal{O}(m^2)$  & $\mathcal{O}(Nm)$  & $\mathcal{O}(m^2)$ & $\mathcal{O}(Nm)$   & $\mathcal{O}(N(d_v{+}1))$ & in-place \\
    \bottomrule
  \end{tabular}}
\end{table}
We form landmarks by average pooling over non-overlapping windows as in~\cite{liu2021swin}. For numerical stability, we replace PyTorch's pseudoinverse with the iterative method of~\cite{razavi2014new}. Steps~2--3 evaluate exponential kernels, which may overflow and destabilize the Nystr\"{o}m computation. We adopt a softmax-style row-max stabilization by introducing diagonal matrices
$\M_L\in\mathbb{R}^{N\times N}$ and $\M_U\in\mathbb{R}^{m\times m}$ as 
\begin{equation*}
\begin{aligned}
(\M_L)_{ii} &= \exp(m_L(i)), &\quad m_L(i) &:= \max_j\Big(\frac{\Q\bar{\K}^{\top}}{\sqrt d}\Big)_{ij}.\\
(\M_U)_{ii} &= \exp(m_U(i)), &\quad m_U(i) &:= \max_j\Big(\frac{\bar{\Q}\K^{\top}}{\sqrt d}\Big)_{ij}.
\end{aligned}
\label{eq:rowmax_scaling}
\end{equation*}
This rescales each row of $\G_L$ and $\G_U$ by its maximum log-score. 
Using $\M_L\M_L^{-1}=\I$ and $\M_U\M_U^{-1}=\I$, we rewrite
\begin{equation*}
\widehat{\G}
=\M_L\M_L^{-1}\G_L\,\G_A^\dagger\,\M_U\M_U^{-1}\G_U
=\M_L\,\G_L^n\,(\G_A^n)^\dagger\,\G_U^n,
\label{eq:nystrom_stable}
\end{equation*}
with normalized blocks
\begin{equation*}
\G_L^n:=\M_L^{-1}\G_L,\quad
\G_A^n:=\M_U^{-1}\G_A,\quad
\G_U^n:=\M_U^{-1}\G_U.
\label{eq:normalized_blocks}
\end{equation*}
Equivalently, $\widehat{\G}=\M_L\widehat{\G}^n$, where $\widehat{\G}^n:=\G_L^n(\G_A^n)^\dagger\G_U^n$ , and the stabilized output in~\eqref{eq:compact_approxattention} is given as
\begin{equation}
\widehat{\boldO}=(\M_L\widehat{\G}^n\V)\oslash(\M_L\widehat{\G}^n\mathbf{1}_N) = (\widehat{\G}^n\V)\oslash(\widehat{\G}^n\mathbf{1}_N).
\label{eq:stable_output}
\end{equation}
In practice, Algorithm~\ref{alg:pnpnystra} computes $\widehat{\boldO}$ using the stabilized form in~\eqref{eq:stable_output} instead of~\eqref{eq:compact_approxattention}. 

\begin{figure*}[!ht]
  \centering
  % -------- Row 1 --------
  \subfloat[Hazy]{%
    \includegraphics[width=0.19\textwidth]{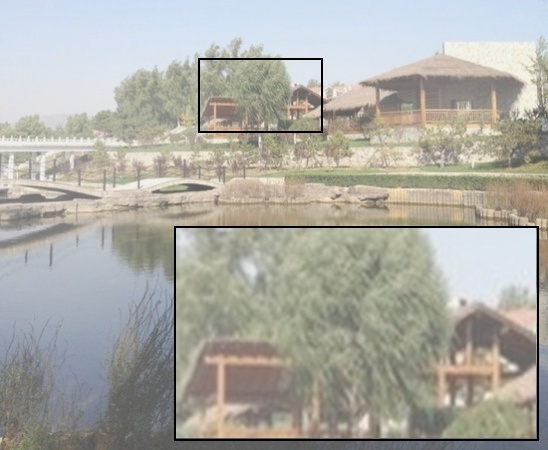}%
  } \hspace{0.03mm}
  \subfloat[\scriptsize{Pretrained} ($27.48$, $0.9582$)]{%
    \includegraphics[width=0.19\textwidth]{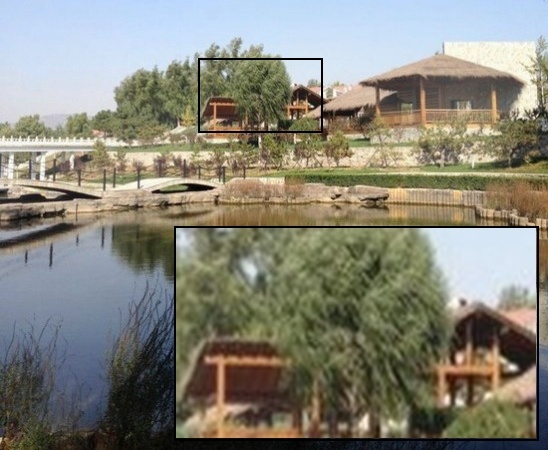}%
  } \hspace{0.03mm}
  \subfloat[\scriptsize{Performer} ($23.27$, $0.9397$)]{%
    \includegraphics[width=0.19\textwidth]{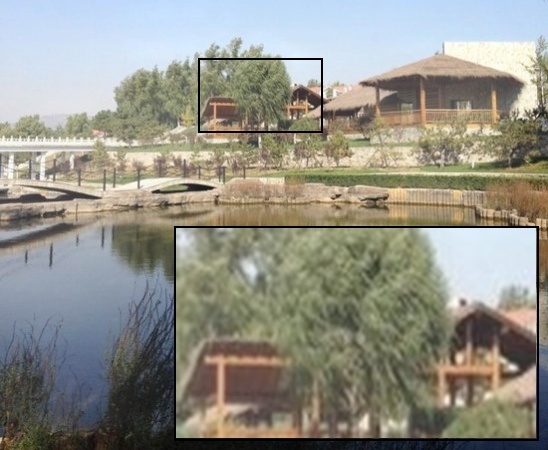}%
  } \hspace{0.03mm}
  \subfloat[\scriptsize{Nystform. ($23.66$, $0.9454$)}]{%
    \includegraphics[width=0.19\textwidth]{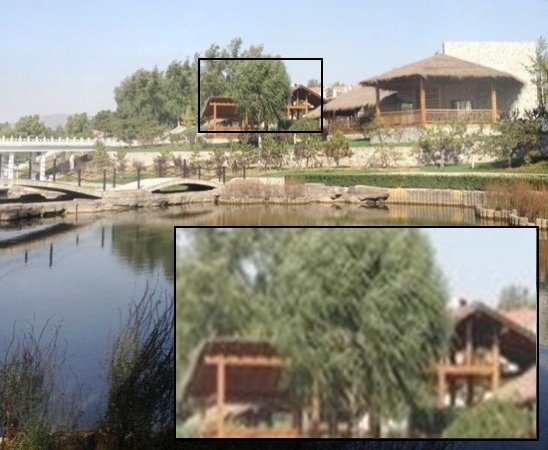}%
  } \hspace{0.03mm}
  \subfloat[\scriptsize{PnP-Nystra} ($\textbf{27.30}$, $\textbf{0.9607}$)]{%
    \includegraphics[width=0.19\textwidth]{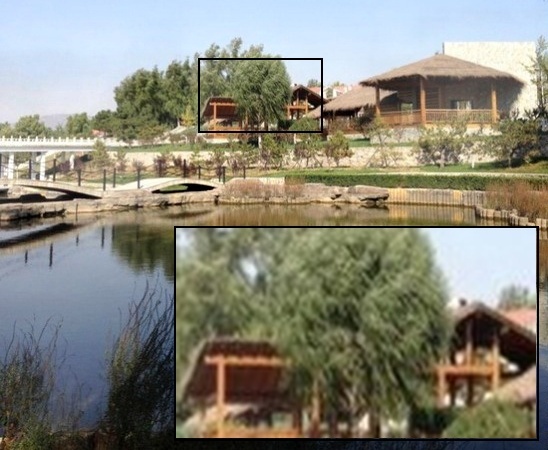}%
  } 

  \vspace{2mm}

  % -------- Row 2 --------
  \subfloat[GT]{%
    \includegraphics[width=0.19\textwidth]{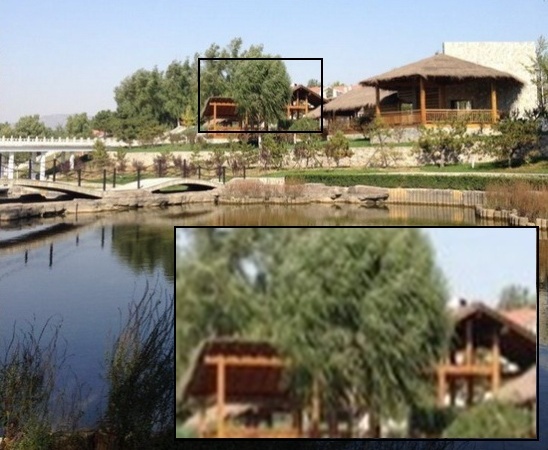}%
  } \hspace{0.03mm}
  \subfloat[\scriptsize{Error: Pretrained - GT}]{%
    \includegraphics[width=0.19\textwidth]{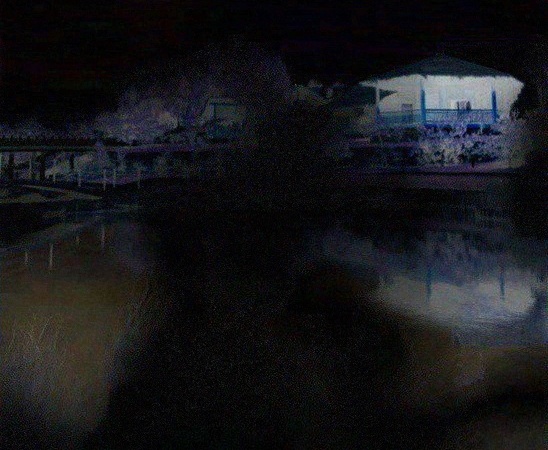}%
  } \hspace{0.03mm}
   \subfloat[\scriptsize{Error: Performer - GT}]{%
    \includegraphics[width=0.19\textwidth]{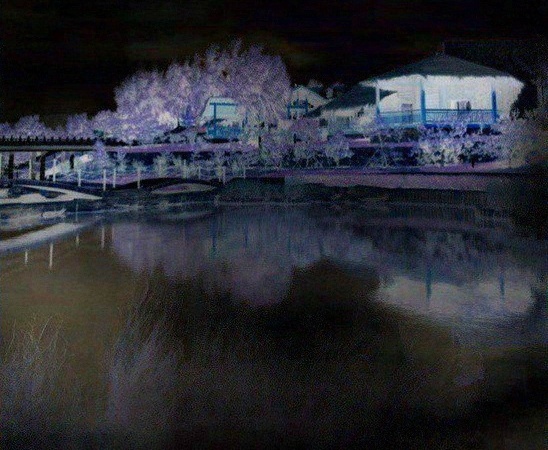}%
  } \hspace{0.03mm}
  \subfloat[\scriptsize{Error: Nystform. - GT}]{%
    \includegraphics[width=0.19\textwidth]{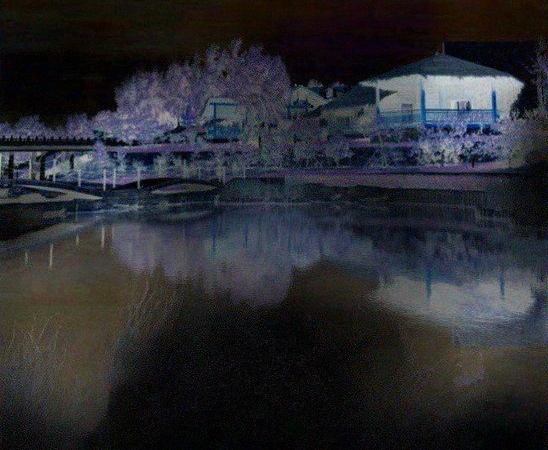}%
  } \hspace{0.03mm}
  \subfloat[\scriptsize{Error: PnP-Nystra - GT}]{%
    \includegraphics[width=0.19\textwidth]{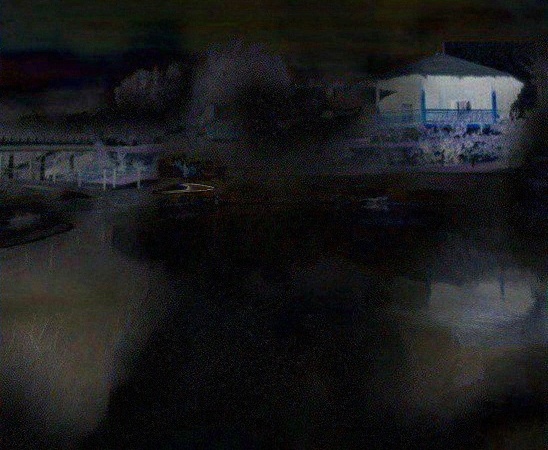}%
  }
\caption{Single-image dehazing. Top: hazy input and restored outputs. Bottom: ground truth and pixel-wise error maps (scaled to [0--255]). PnP-Nystra matches the pretrained model closely while producing cleaner dehazing than other training-free linear attention baselines, as evidenced by metrics (PSNR(dB), SSIM) and reconstruction of trees and houses in the zoomed area.}

  \label{fig:qualitative_dehaze}
\end{figure*}

\section{Experiments}
\label{sec:experiments}

We evaluate drop-in replacement capabilities of {PnP-Nystra} in existing restoration models, followed by an ablation study of its algorithmic parameters. PSNR and SSIM metrics for the experiments are reported, alongside runtime and speedup on both GPU and CPU platforms. All timings are averaged over multiple inference runs per test dataset. Experiments are conducted using PyTorch $2.7$ with CUDA $12.2$, with GPU and CPU benchmarks performed on NVIDIA RTX $4090$ and Intel Xeon Gold processors, respectively. 
The implementation is available at: \url{https://anonymous.4open.science/r/PnP_Nystra-0F31}.

\subsection{PnP-Nystra for restoration}
In this section, we evaluate \textbf{PnP-Nystra} as a drop-in replacement for MHSA in three state-of-the-art restoration Transformers: \textbf{SwinIR}~\cite{swinir}, \textbf{Uformer-B}~\cite{uformer}, and \textbf{Dehazeformer}~\cite{song2023vision}. We keep all pretrained projection weights $(\mathbf{W}_Q,\mathbf{W}_K,\mathbf{W}_V)$ fixed and perform no finetuning. Unless stated otherwise, we use {$m=16$} landmarks and {$4$--$6$} pseudoinverse iterations, chosen to provide strong performance across tasks. We compare against other training-free linear attention substitutes, such as Performer~\cite{choromanski2020rethinking} and Nystr\"{o}mformer~\cite{xiong2021nystromformer}. Our intent is not exhaustive restoration benchmarking, but to quantify the inference-time trade-off of attention approximation within these high-performing pretrained baselines. Speedups over the original attention are summarized in Table~\ref{tab:timing_all}.

\begin{table}[!htp]
\centering
\caption{Timing comparison per attention block for  original model and PnP-Nystra. {GPU/CPU} time (in ms) is reported. }
\label{tab:timing_all}
\scriptsize
\setlength{\tabcolsep}{4pt}
%\resizebox{\textwidth}{!}{%
\begin{tabular}{l c c c c}
\toprule
\textbf{Resolution} & \textbf{Scale} & \textbf{Original}  & \textbf{PnP-Nystra} & \textbf{Speedup} \\
\midrule

\multicolumn{5}{c}{\textbf{SwinIR for Super-Resolution }}\\
\midrule
$288\times 288$ & $2$ & $524.61$ / $17916$ & $235.76$ / $8139$ & $2.23\times$ / $2.20\times$ \\
${160\times 160}$ & ${4}$ & $159.96$ / $5298$  & $65.46$  / $1688$ & $2.44\times$ / $3.13\times$ \\
$96\times 96$   & $8$ & $59.18$  / $2274$  & $29.11$  / $652$  & $2.03\times$ / $3.49\times$ \\
\midrule
\multicolumn{5}{c}{\textbf{Uformer-B for Denoising }}\\
\midrule
$512\times 512$ & -- & $218.01$ / $6958$ & $67.26$ / $1853$ & $3.24\times$ / $3.75\times$ \\
$256\times 256$ & -- & $81.81$  / $2610$ & $22.64$ / $372$  & $3.61\times$ / $7.05\times$ \\
\midrule

\multicolumn{5}{c}{\textbf{Uformer-B for Deblurring}}\\
\midrule
${768\times 768}$ & -- & $285.46$ / $9098$ & $155.19$ / $4942$ & $1.84\times$ / $1.84\times$ \\
\midrule

\multicolumn{5}{c}{\textbf{Dehazeformer for Dehazing}}\\
\midrule
% Replace the duplicate rows below with your actual dehazing measurements if different
$416\times 576$ & -- & $143.92$ / $5106$ & $51.71$ / $1689$ & $2.78\times$ / $3.02\times$ \\
$576\times 576$ & -- & $183.53$ / $6938$ & $70.05$ / $2558$ & $2.62\times$ / $2.71\times$ \\
$736\times 576$ & -- & $261.80$ / $10192$ & $105.02$ / $4157$ & $2.49\times$ / $2.45\times$ \\
\bottomrule
\end{tabular}%
%}
\end{table}

\subsubsection{Dehazing}
We evaluate \textbf{Dehazeformer}~\cite{song2023vision} with \textbf{PnP-Nystra} as a drop-in replacement for MHSA on single-image dehazing. We use a window size of $32$ (i.e., $N=32^2$ tokens) and run $5$ pseudoinverse iterations for all dehazing  experiments. Table~\ref{tab:dehaze_accuracy} reports PSNR/SSIM on different datasets, where PnP-Nystra consistently outperforms other training-free linear attention substitutes, producing accuracy close to (and in some cases exceeding) the original pretrained Dehazeformer. This is achieved at a $2.5 \times$ to $2.8 \times$ speedup as shown in the Table.~\ref{tab:timing_all}. Qualitative results in Fig.~\ref{fig:qualitative_dehaze} show that PnP-Nystra preserves fine details and contrast on par with the original model, while producing cleaner haze removal than other training-free approximations.

\begin{table}[!htp]
  \centering
  \caption{Accuracy Comparison for Image Dehazing (PSNR / SSIM) with other linear attention methods}
  \label{tab:dehaze_accuracy}
  \renewcommand{\arraystretch}{1.2}
  \setlength{\tabcolsep}{8pt}
  \resizebox{\columnwidth}{!}{%
    \begin{tabular}{l| c| c c c}
      \toprule
      \textbf{Dataset}
        & \textbf{Original}
        & \textbf{Performer}
        & \textbf{Nystromformer}
        & \textbf{PnP-Nystra} \\
      \midrule
      Reside-Out 
        & $26.51$ / $0.9512$
        & $25.26$ / $0.9328$
        & $23.57$ / $0.9266$
        & $\textbf{26.34}$ / $\textbf{0.9510}$ \\
      RSHaze
        & $32.80$ / $0.9261$
        & $29.00$ / $0.8957$
        & $33.18$ / $0.9363$
        & $\textbf{33.58}$ / $\textbf{0.9405}$ \\
      RESIDE-6K
        & $24.79$ / $0.9352$
        & $21.00$ / $0.8834$
        & $21.37$ / $0.8962$
        & $\textbf{23.08}$ / $\textbf{0.9163}$ \\

      \bottomrule
    \end{tabular}%
  }
\end{table}

\subsubsection{Denoising and Deblurring}
We evaluate the \textbf{Uformer-B} model with \textbf{PnP-Nystra} on both image denoising (SIDD, BSDS200) and image deblurring (RealBlur-R). For denoising, we use a window size $64$ (i.e., $N=64^2$ tokens) and report results at two input resolutions in Table~\ref{tab:acc_denoise_deblur} and runtime in Table~\ref{tab:timing_all}. PnP-Nystra achieves $2.0$--$2.7\times$ GPU and $3.6$--$7\times$ CPU speedups, with the smallest accuracy drop among training-free linear attention baselines (PSNR drop of $\sim 1$~dB and SSIM decrease of $\sim 0.02$)
. For deblurring on RealBlur-R, we use window size $32$ (i.e., $N=32^2$) and obtain $1.84\times$ GPU and CPU speedups (Table~\ref{tab:timing_all}) with negligible change in PSNR/SSIM (Table~\ref{tab:acc_denoise_deblur}).

\begin{table}[!htp]
  \centering
  \caption{Accuracy comparison (PSNR / SSIM) for Uformer-B with training-free linear attention substitutes.}
  \label{tab:acc_denoise_deblur}
  \renewcommand{\arraystretch}{1.15}
  \setlength{\tabcolsep}{6pt}
  \resizebox{\columnwidth}{!}{%
    \begin{tabular}{l|c|c c c}
      \toprule
      \textbf{Dataset} &
      \textbf{Original} &
      \textbf{Performer} &
      \textbf{Nystr\"{o}mformer} &
      \textbf{PnP-Nystra} \\
      \midrule

      \multicolumn{5}{c}{\textbf{Denoising} } \\
      \midrule
      SIDD     & $38.89$ / $0.8950$ & $34.80$ / $0.8713$ & $37.46$ / $0.8838$ & $\mathbf{37.88}$ / $\mathbf{0.8842}$ \\
      BSDS200  & $28.05$ / $0.8067$ & $24.80$ / $0.7469$ & $26.77$ / $0.7829$ & $\mathbf{27.17}$ / $\mathbf{0.7838}$ \\
      \midrule

      \multicolumn{5}{c}{\textbf{Deblurring}} \\
      \midrule
      RealBlur-R & $33.98$ / $0.9404$ & $33.97$ / $0.9384$ & $34.01$ / $0.9407$ & $\mathbf{34.02}$ / $\mathbf{0.9407}$ \\
      \bottomrule
    \end{tabular}%
  }
\end{table}

\subsubsection{Super-resolution}
Table~\ref{tab:sr_accuracy} compares SwinIR and its PnP-Nystra variant on Set5 and BSDS100 across different scales for a window size of $32$, i.e., $N = 32^2$. As shown in the Table.~\ref{tab:timing_all}, PnP-Nystra consistently achieves $2$-$3\times$ speedup on both GPU and CPU, with better runtime benefits at lower resolutions. Even with the speedup, the accuracy drop, measured in PSNR and SSIM, is lowest among the compared linear attention methods in most evaluations in the Table.~\ref{tab:sr_accuracy} across different scaling factors. 
% We also demonstrate in Fig.~\ref{fig:qualitative_superres} that PnP-Nystra achieves super-resolution results visually on par with the original SwinIR, with negligible perceptual difference. 

\begin{table}[!ht]
  \centering
  \caption{SwinIR super-resolution accuracy (PSNR(dB),SSIM) comparison across scaling factors.}

  \label{tab:sr_accuracy}
  \renewcommand{\arraystretch}{1.3}
  \resizebox{\columnwidth}{!}{%
    \begin{tabular}{c c| c| c c c }
      \toprule
      \textbf{Dataset} & \textbf{Scale}
        & \textbf{Original}
        & \textbf{Performer}
        & \textbf{Nystr\"{o}mformer}
        & \textbf{PnP-Nystra} \\
      \midrule
      \multirow{3}{*}{Set5}
        & $2$ & $36.13$ / $0.9455$ 
        & $33.91$ / $0.9322$ & $35.15$ / $0.9413$ & $\textbf{35.25}$ / $\textbf{0.9416}$ \\
        & $4$ & $30.86$ / $0.8729$ 
        & $28.11$ / $0.8142$ & $\textbf{29.26}$ / $\textbf{0.8467}$ & ${29.10}$ / ${0.8437}$ \\
        & $8$ & $25.62$ / $0.7439$ 
        & $23.53$ / $0.6434$ & $24.26$ / $0.6787$ & $\textbf{24.41}$ / $\textbf{0.6853}$ \\
      \midrule
      \multirow{3}{*}{BSDS100}
        & $2$ & $30.98$ / $0.8922$&
        $29.82$ / $0.8740$ & $30.64$ / $0.8821$ & $\textbf{30.71}$ / $\textbf{0.8830}$ \\
        & $4$ & $26.43$ / $0.7183$ 
        & $24.98$ / $0.6815$ & $25.81$ / $0.6992$ & $\textbf{25.92}$ / $\textbf{0.7008}$ \\
        & $8$ & $23.61$ / $0.5680$& 
        $22.73$ / $0.5291$ & $23.21$ / $0.5487$ & $\textbf{23.29}$ / $\textbf{0.5503}$ \\
      \bottomrule
    \end{tabular}%
  }
\end{table}

\subsection{Ablation Study of PnP-Nystra}
In this section, we present an ablation study of {PnP-Nystra}, with emphasis on the internal hyperparameters.

\begin{figure}[!ht]
  \centering
  \begin{subfigure}[b]{0.47\columnwidth}
    \centering
    \includegraphics[width=\textwidth]{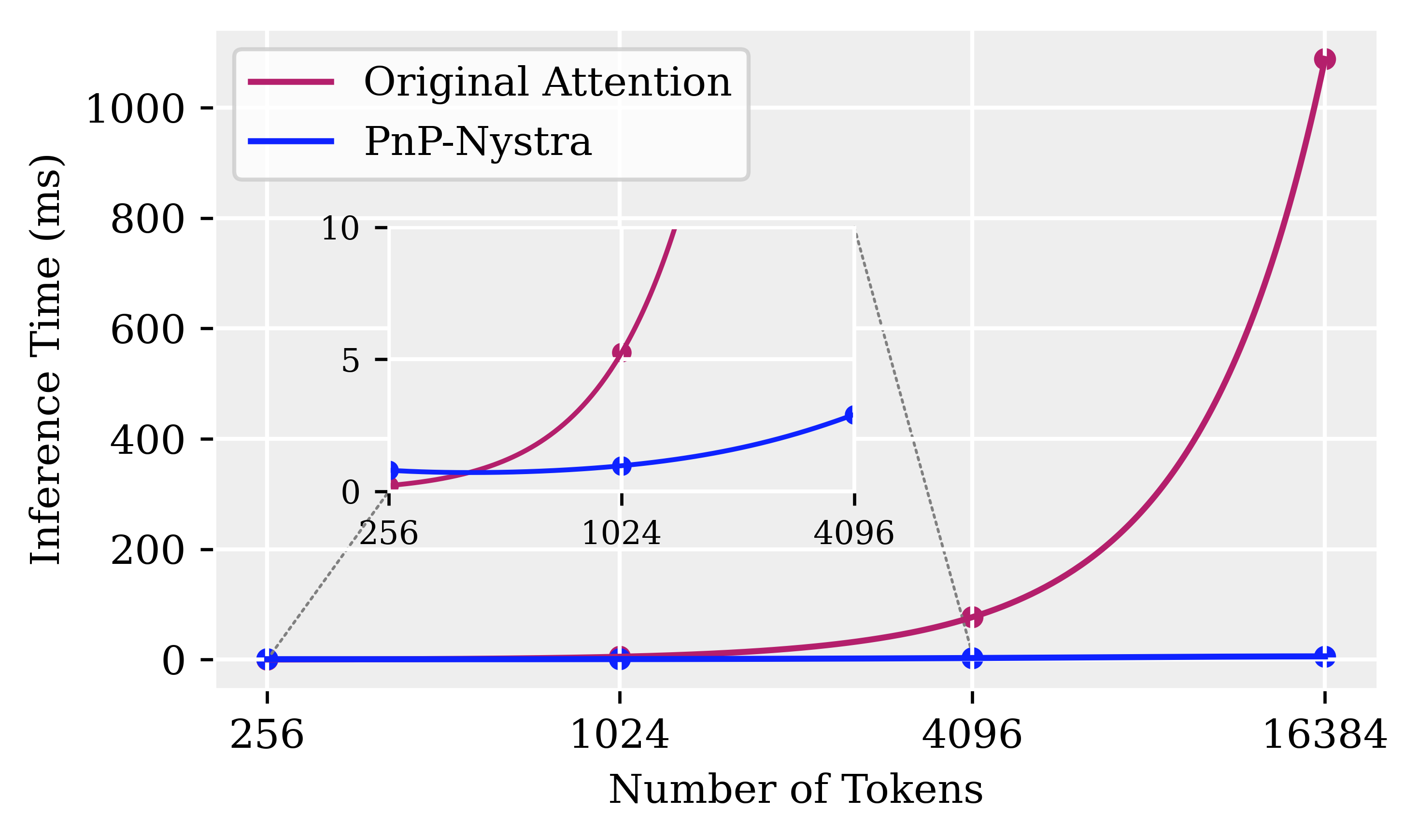}
    \caption{Linear scale}
    \label{fig:linear}
  \end{subfigure}\hfill
  \begin{subfigure}[b]{0.47\columnwidth}
    \centering
    \includegraphics[width=\textwidth]{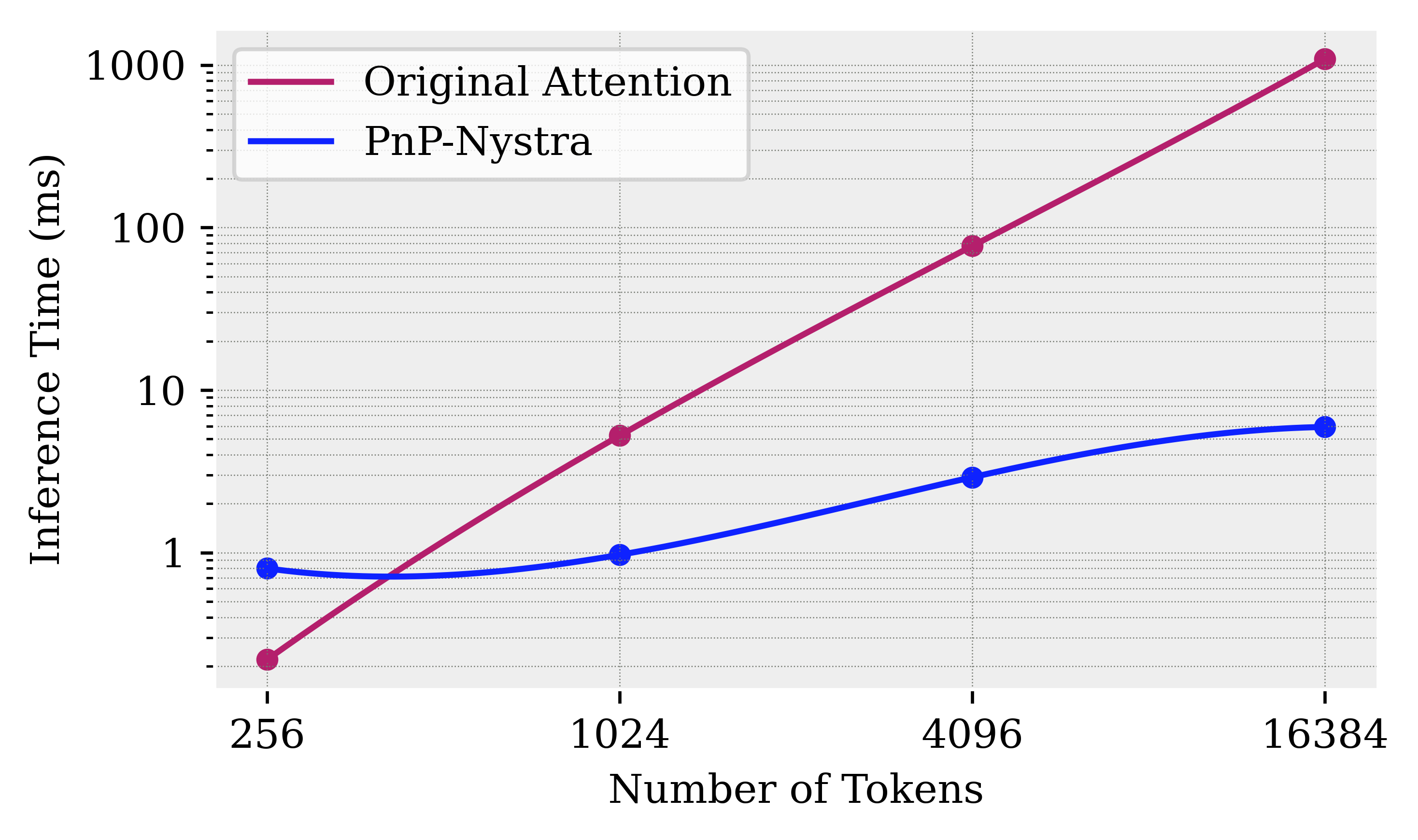}
    \caption{Logarithmic scale}
    \label{fig:log}
  \end{subfigure}
 \caption{Inference time vs. token count $N$: Unlike MHSA which grows quadratically with $N$, PnP-Nystra scales linearly.}

  \label{fig:scale_comparison}
\end{figure}

\noindent \textbf{Scaling of Inference Time with $N$:} Fig.~\ref{fig:scale_comparison} presents the CPU inference runtime of the original MHSA module versus {PnP-Nystra} (averaged over repeated runs), plotted in both linear and logarithmic scales with respect to the token count~$N$. In line with the complexity analysis, {PnP-Nystra} exhibits linear scaling with an increasing number of tokens.

\noindent \textbf{Landmarks and Pseudoinverse iterations:} We evaluate the effect of varying the number of landmarks \( m \) and the number of iterations used for pseudoinverse computation in Step $4$ of Algorithm \ref{alg:pnpnystra}.  Table~\ref{tab:psnr_ssim_diff_ablations} reports PSNR and SSIM drop for the PnP-Nystra variant of Uformer-B for denoising on the SIDD dataset. As expected from  Nystr\"{o}m approximation, the approximation performance tends to the original with an increase in the number of landmarks. Additionally, a better estimate of the pseudoinverse leads to improved approximation. 

\begin{table}[!ht]
  \centering
  \caption{Impact of Number of Landmarks and Pseudoinverse Iterations on Approximation Accuracy}
  \label{tab:psnr_ssim_diff_ablations}
  \resizebox{0.95\columnwidth}{!}{%
  \begin{tabular}{@{} c c c @{\hskip 1.5em} c c c @{}}
    \toprule
    \multicolumn{3}{c}{\textbf{Varying \# Landmarks (at 6 Iterations)}} 
    & \multicolumn{3}{c}{\textbf{Varying Iterations (at 16 Landmarks)}} \\
    \cmidrule(lr){1-3} \cmidrule(l){4-6}
    \# Landmarks & PSNR Drop & SSIM Drop 
    & Iterations & PSNR Drop & SSIM Drop \\
    \midrule   
     $8$  & $1.54$ & $0.0127$  & $1$ & $2.84$ & $0.0167$ \\
    $16$  & $1.01$ & $0.0108$  & $3$ & $1.94$ & $0.0138$ \\
    $32$  & $0.86$ & $0.0102$  & $6$ & $1.01$ & $0.0108$ \\
    \bottomrule
  \end{tabular}
  }
\end{table}

\begin{figure}[!ht]
  \centering

  \vspace{2mm}
  \setcounter{subfigure}{0} 
  \subfloat[Original Model]{%
    \includegraphics[width=0.45\columnwidth]{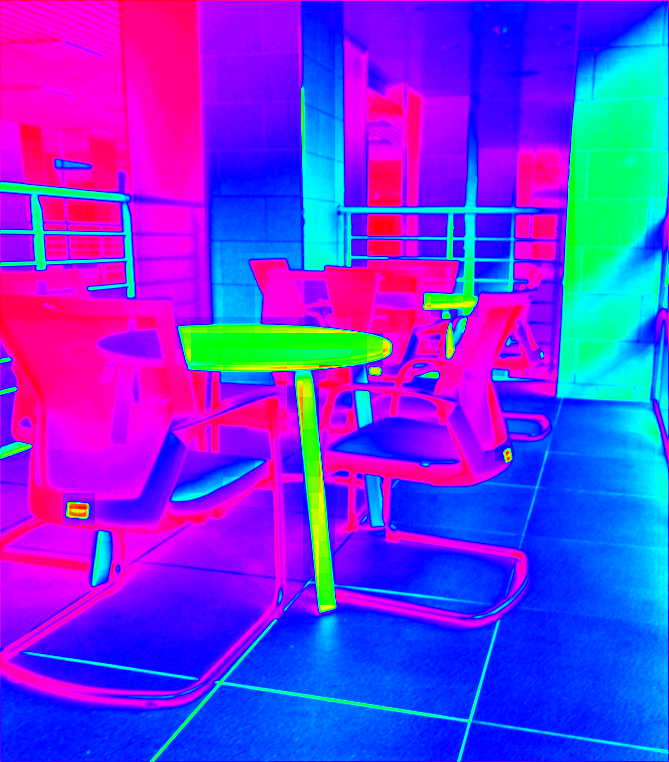}%
  }\hspace{2mm}
    \subfloat[PnP-Nystra]{%
    \includegraphics[width=0.45\columnwidth]{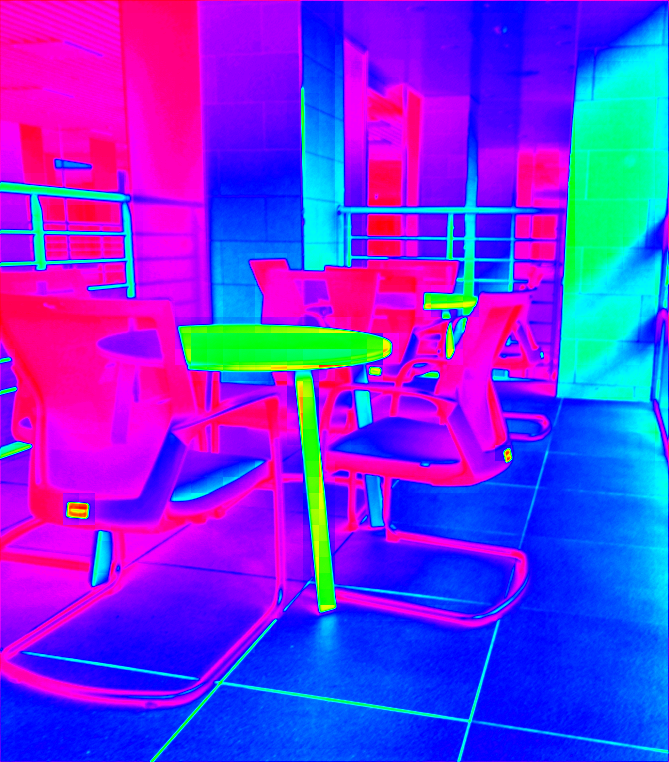}%
  }
\caption{Attention map visualization for Uformer-B: (a) original model and (b) PnP-Nystra. Both maps have the same salient structures with strong responses along object boundaries}

 \label{fig:attention}
\end{figure}

\noindent\textit{\textbf{Attention map comparison:}}
PnP-Nystra approximates the kernel matrix $\G$ and therefore induces an approximate attention map $\S$. As shown in Fig.~\ref{fig:attention}, the resulting attention distribution closely follows that of the original model, highlighting the same prominent edges and structural regions while maintaining similar suppression over less informative areas.

\section{Conclusion and Future Work}
We presented PnP-Nystra, a plug-and-play Nystr\"{o}m approximation for self-attention that enables efficient inference in window-based transformer models for image  restoration. Experiments show that PnP-Nystra acts as a fast, training-free replacement for self-attention in pre-trained models, offering a viable solution for resource-constrained deployment. Future work will focus on extending PnP-Nystra to global attention in vision transformers and diffusion transformers, and  spatiotemporal attention for video restoration.

\bibliographystyle{IEEEbib}
\bibliography{myrefs}

@article{attn,
title={Attention is all you need},
  author={Vaswani, Ashish and Shazeer, Noam and Parmar, Niki and Uszkoreit, Jakob and Jones, Llion and Gomez, Aidan N and Kaiser, {\L}ukasz and Polosukhin, Illia},
  journal={Proc. Adv. Neural Inf. Process. Syst.},
  volume={30},
  year={2017}
}

@article{nemtsov2016matrix,
  author  = {A. Nemtsov, A. Averbuch and A. Schclar},
  title   = {Matrix compression using the {NystrÃ¶m} method},
  journal = {Intell. Data Anal.},
  year    = {2016},
  volume  = {20},
  number  = {5},
  pages   = {997--1019},
}

@article{convnet,
 title={A convnet for the 2020s},
  author={Liu, Zhuang and Mao, Hanzi and Wu, Chao-Yuan and Feichtenhofer, Christoph and Darrell, Trevor and Xie, Saining},
  journal={Proc. IEEE/CVF Conf. Comput. Vis. Pattern Recognit.},
  pages={11976--11986},
  year={2022}
}

@article{ren2024ninth,
  title={The ninth {NTIRE} 2024 efficient super-resolution challenge report},
  author={Ren, Bin and Li, Yawei and Mehta, Nancy and Timofte, Radu and Yu, Hongyuan and Wan, Cheng and Hong, Yuxin and Han, Bingnan and Wu, Zhuoyuan and Zou, Yajun and others},
  journal={Proc. IEEE/CVF Conf. Comput. Vis. Pattern Recognit.},
  pages={6595--6631},
  year={2024}
}

@article{nystrom1930praktische,
  author    = {E. J. Nystr{\"o}m},
  title     = {{\"U}ber die praktische Aufl{\"o}sung von Integralgleichungen mit Anwendungen auf Randwertaufgaben},
  journal = {Acta Math.},
  volume    = {54},
  number    = {1},
  pages     = {185--204},
  year      = {1930}
}

@book{baker1977numerical,
  author    = {C. T. Baker},
  title     = {The Numerical Treatment of Integral Equations},
  publisher = {Clarendon Press},
  year      = {1977}
}

@article{talebi2014global,
  author    = {H. Talebi and P. Milanfar},
  title     = {Global Image Denoising},
  journal   = {IEEE Trans. Image Process.},
  volume    = {23},
  number    = {2},
  pages     = {755--768},
  year      = {2014}
}

@article{williams2000using,
  title={Using the {N}ystr{\"o}m method to speed up kernel machines},
  author={Williams, Christopher and Seeger, Matthias},
  journal={Proc. Adv. Neural Inf. Process. Syst.},
  volume={13},
  year={2000}
}

@article{fowlkes2004spectral,
  author    = {C. Fowlkes and S. Belongie and F. Chung and J. Malik},
  title     = {Spectral Grouping Using the {Nystr\"{o}m} Method},
  journal = {IEEE Trans. Pattern Anal. Mach. Intell.},
  volume    = {26},
  number    = {2},
  pages     = {214--225},
  year      = {2004}
}

@article{swinir,
  title={Swinir: {Image} restoration using swin transformer},
  author={Liang, Jingyun and Cao, Jiezhang and Sun, Guolei and Zhang, Kai and Van Gool, Luc and Timofte, Radu},
  journal={Proc. IEEE Int. Conf. Comput. Vis.},
  pages={1833--1844},
  year={2021}
}

@article{uformer,
  title={Uformer: {A} general u-shaped transformer for image restoration},
  author={Wang, Zhendong and Cun, Xiaodong and Bao, Jianmin and Zhou, Wengang and Liu, Jianzhuang and Li, Houqiang},
  journal={Proc. IEEE/CVF Conf. Comput. Vis. Pattern Recognit.},
  pages={17683--17693},
  year={2022}
}

@article{linformer,
  title={Linformer: Self-attention with linear complexity},
  author={Wang, Sinong and Li, Belinda Z and Khabsa, Madian and Fang, Han and Ma, Hao},
  journal={arXiv preprint arXiv:2006.04768},
  year={2020}
}

@article{flashattention,
   title={Flashattention: Fast and memory-efficient exact attention with io-awareness},
  author={Dao, Tri and Fu, Dan and Ermon, Stefano and Rudra, Atri and R{\'e}, Christopher},
  journal={Proc. Adv. Neural Inf. Process. Syst.},
  volume={35},
  pages={16344--16359},
  year={2022}
}

@article{sparseformer,
  title={Sparseformer: Sparse visual recognition via limited latent tokens},
  author={Gao, Ziteng and Tong, Zhan and Wang, Limin and Shou, Mike Zheng},
  journal={Proc. IEEE Int. Conf. Comput. Vis.},
  year={2024}
}

@article{tillet2019triton,
  title={Triton: An intermediate language and compiler for tiled neural network computations},
  author={Tillet, Philippe and Kung, Hsiang-Tsung and Cox, David},
  journal= {Proc. ACM SIGPLAN Int. Workshop Mach. Learn. Program. Lang.},
  pages={10--19},
  year={2019}
}

@article{mahoney2009cur,
  title={{CUR} matrix decompositions for improved data analysis},
  author={Mahoney, Michael W and Drineas, Petros},
  journal = {Proc. Natl. Acad. Sci. U.S.A.},
  volume={106},
  number={3},
  pages={697--702},
  year={2009},
  publisher={National Academy of Sciences}
}

@article{liu2021swin,
  title={Swin transformer: Hierarchical vision transformer using shifted windows},
  author={Liu, Ze and Lin, Yutong and Cao, Yue and Hu, Han and Wei, Yixuan and Zhang, Zheng and Lin, Stephen and Guo, Baining},
  journal={Proc. IEEE Int. Conf. Comput. Vis.},
  pages={10012--10022},
  year={2021}
}

@article{wang2013improving,
  title={Improving {CUR} matrix decomposition and the {N}ystr{\"o}m approximation via adaptive sampling},
  author={Wang, Shusen and Zhang, Zhihua},
  journal = {J. Mach. Learn. Res.},
  volume={14},
  number={1},
  pages={2729--2769},
  year={2013},
}

@article{razavi2014new,
 title={A new iterative method for finding approximate inverses of complex matrices},
  author={Razavi, M Kafaei and Kerayechian, Asghar and Gachpazan, Mortaza and Shateyi, Stanford},
  journal = {Abstr. Appl. Anal.},
  volume={2014},
  number={1},
  pages={563787},
  year={2014},
}

@article{choromanski2020rethinking,
  title={Rethinking attention with performers},
  author={Choromanski, Krzysztof and Likhosherstov, Valerii and Dohan, David and Song, Xingyou and Gane, Andreea and Sarlos, Tamas and Hawkins, Peter and Davis, Jared and Mohiuddin, Afroz and Kaiser, Lukasz and others},
  journal={Proc. Int. Conf. Learn. Represent.},
  year={2021}
}

@article{xiong2021nystromformer,
  title={Nystr{\"o}mformer: A nystr{\"o}m-based algorithm for approximating self-attention},
  author={Xiong, Yunyang and Zeng, Zhanpeng and Chakraborty, Rudrasis and Tan, Mingxing and Fung, Glenn and Li, Yin and Singh, Vikas},
  journal={Proc. AAAI Conf. Artif. Intell.},
  volume={35},
  number={16},
  pages={14138--14148},
  year={2021}
}

@article{ho2019axial,
  title={Axial attention in multidimensional transformers},
  author={Ho, Jonathan and Kalchbrenner, Nal and Weissenborn, Dirk and Salimans, Tim},
  journal={arXiv preprint arXiv:1912.12180},
  year={2019}
}

@article{song2023vision,
  title={Vision transformers for single image dehazing},
  author={Song, Yuda and He, Zhuqing and Qian, Hui and Du, Xin},
  journal={IEEE Trans. Image Process.},
  volume={32},
  pages={1927--1941},
  year={2023}
}
\end{document}